\documentclass[12pt,letterpaper]{JHEP3}

\usepackage{amsmath}
\usepackage{cite}

\def\pa{\partial}

\def\to{\rightarrow}
\def\be{\begin{equation}}
\def\ee{\end{equation}}
\def\bea{\begin{eqnarray}}
\def\eea{\end{eqnarray}}
\def\nonu{\nonumber \\{}}
\def\half{{1 \over 2}}
\def\cq{{\cal{Q}}}

\def\a{\alpha}
\def\b{\beta}

\def\d{\delta}
\def\e{\epsilon}
\def\f{\phi}

\def\h{\eta}

\def\l{\lambda}
\def\m{\mu}
\def\n{\nu}
\def\o{\omega}
\def\p{\pi}

\def\x{\xi}

\title{Superconformal Quantum Mechanics of Small Black Holes}

\author{\centerline{Seok Kim
and Joris Raeymaekers}\\

\centerline{School of Physics, Korea Institute for
Advanced Study,}

\centerline{207-43, Cheongnyangni 2-Dong, Dongdaemun-Gu, Seoul
130-722, Korea}

\bigskip
\centerline{{\rm E-mail}:\email{seok, joris@kias.re.kr}}}

\abstract{
Recently, Gaiotto, Strominger and Yin have proposed a holographic
dual description for the near-horizon
physics of certain $N=2$ black holes in terms of the superconformal quantum mechanics
on D0-branes in the attractor geometry.
We provide further evidence for their proposal by applying it to the case of `small' black holes which have vanishing horizon area in the
leading supergravity approximation. We consider 2-charge black holes in type IIA on $T^2 \times M$, where $M$ can be
either $K_3$ or $T^4$, made up out of D0-branes and D4-branes wrapping  $M$. We construct
the corresponding superconformal quantum mechanics and show that the asymptotic growth of chiral
primaries exactly matches with the known entropy of these black holes. The state-counting problem reduces
to counting lowest Landau levels on $T^2$ and Dolbeault cohomology classes on $M$.}


\preprint{\hepth{0505176} \\ KIAS-P05033}

\begin{document}
\section{Introduction}

Compactifications of type II theory on a Calabi-Yau manifold $CY_3$  contain extremal black holes arising from wrapping
 branes
around cycles of $CY_3$ whose near-horizon region is an $AdS_2 \times S^2 \times CY_3$ attractor geometry
\cite{Ferrara:1995ih,Strominger:1996kf}.
String theory in this background is expected to be holographically dual to a conformal quantum mechanics
 \cite{Maldacena:1997re}, but this $AdS_2/CFT_1$ duality is much less well-understood than its
higher-dimensional counterparts (see however \cite{Michelson:1999zf,Ooguri:2004zv}).

However, for a class of black holes in type IIA on $CY_3$
carrying D0 and
D4-brane charges, a concrete proposal for such a holographic dual $CFT_1$ was proposed by Gaiotto, Strominger
and Yin (GSY)  \cite{Gaiotto:2004ij}.
The $CFT_1$ takes the form of a quantum mechanics of $N$ D0 brane probes moving
in the near-horizon geometry.
The super-isometry group of the background
 acts as a superconformal symmetry group on the quantum mechanics \cite{Claus:1998ts,Gibbons:1998fa}.
 It was proposed that the black hole ground states should be identified with
 the chiral primaries of this quantum mechanical system.
Of particular importance were nonabelian $N$-D0 configurations corresponding to D2-branes
wrapping the black hole horizon and carrying $N$ units of worldvolume magnetic flux \cite{Simons:2004nm,Gaiotto:2004pc}.
These
experience a magnetic flux along the Calabi-Yau directions induced by the D4-branes in the background.
The chiral primary states correspond to lowest Landau levels, and their degeneracy was found to
exactly reproduce the leading order entropy formula. However, the D0-D2 bound states alone did
not correctly account for the known subleading corrections to the entropy formula.

The analysis of GSY was performed for `large' black holes, which have a nonvanishing horizon
area in the leading supergravity approximation. Here the D4-brane charges $p^A$ are restricted
to obey the condition
$$ D \equiv {1 \over 6} C_{ABC} p^A p^B p^C \neq 0 $$ where $C_{ABC}$ are the triple intersection numbers on $CY_3$.
Furthermore, all $p^A$ have to be taken to be nonvanishing and large in order for $\a'$ corrections to the background
to be suppressed.

In the present work we find additional evidence for the GSY proposal by applying it to a different
class of black holes which have vanishing horizon area in the leading supergravity approximation,
but acquire a string scale horizon when higher derivative corrections are included
\cite{Sen:1995in,Sen:1997is,Dabholkar:2004yr,Dabholkar:2004dq}. For these `small' black holes,
the quantity $D$ vanishes.
We will limit ourselves to two special cases where the number of supersymmetries preserved by  the background
is enlarged and where the analysis becomes more tractable. We consider $D0-D4$ black holes in compactifications on
$T^2 \times M$, where $M$ can be either $K_3$ or $T^4$, and where the D4-branes are wrapped on $M$.
These black holes are $1/2$ BPS states in $N=4$ and $1/4$ BPS states in $N=8$ supergravity respectively.
In contrast to case of large black holes, only one of the magnetic charges $p^A$ is nonzero  and
 the D4-brane magnetic flux does not permeate all cycles
in the compactification manifold but  only has a component along $T^2$.
As we shall show, this leads to a modified near-horizon behavior  of the K\"ahler moduli
of $M$. Placing a horizon-wrapping D2 brane with D0-brane flux in this background, we will find a quantum
mechanics with a symmetry
algebra that is a direct sum of an $N=4$ superconformal algebra and the algebra of $N=4$ supersymmetric
quantum mechanics  on $M$. The counting of chiral primaries now reduces to counting lowest Landau levels
on $T^2$ and Dolbeault cohomology classes on $M$. Their asymptotic degeneracy is found to reproduce exactly the
leading order entropy formula in both cases.

\section{Quantum mechanics of the 2-charge black hole on $K_3 \times T^2$}

\subsection{Near-horizon geometry}

We consider type IIA compactified on $K_3 \times T^2$ in the
presence of  D0-branes and  D4-branes wrapped on the
$K_3$. We choose a basis $\{\o_A\}_{A = 1 \ldots 23}$ of 2-forms
on $T^2 \times K_3$ in which $\o_1$ is the volume form on $T^2$
and $\{\o_i\}_{i = 2 \ldots 23}$ are the 2-forms on $K_3$.
The 4-dimensional effective theory is an $N=4$ supergravity theory but we shall work in
the $N=2$ formalism of \cite{deWit:1979ug}.
It contains 24 homogeneous complex scalars $X^I,\ I= 0 \dots 23$
and corresponding $U(1)$ gauge fields $F^I_{\m \n}$. Electric and magnetic
charges are labelled by integers $(q_I, p^I)$. The D0-D4 system of interest
carries nonzero $q_0$ and $p^1$, with all other charges set to zero.

The prepotential, including the leading quantum correction, is
given by
$$
F = - \half C_{ij} X^i X^j {X^1 \over X^0} - {1 \over 64} \hat{A} {X^1 \over X^0}
$$
where $C_{ij} = \int_{K_3} \o_i \wedge \o_j$ is the intersection matrix on $K_3$ and $\hat{A}$ is the
square of the graviphoton field strength.

As in \cite{Sen:2004dp}, we impose a reality condition such that $X^0$ is real and the $X^A$ are
imaginary, so that $F$ is also imaginary. The equations of motion for
a static, spherically symmetric BPS solution carrying charges $(q_I,p^I)$ then reduce to
\cite{LopesCardoso:2000qm,Sen:2004dp}:
\be
\begin{array}{c}
 ds^2 = - e^{2 g(r)} dt^2 + e^{- 2 g(r)} d \vec{x}^2; \qquad r \equiv \sqrt{ \vec{x}^2} \\
 e^{-g} ( X^I - \bar X^I ) = i ( h^I + {p^I \over r}); \qquad
 e^{-g} ( F_I - \bar F_I ) = i ( h_I + {q_I \over r}) \\
\hat{A} = - 64 e^{2 g} (g')^2; \qquad
 e^{-K} + \half \chi = - 128 i e^{3 g} {1 \over r^2} \pa_r \left( r^2 e^{-g} g'(F_{\hat{A}} -
\bar F_{\hat{A}}) \right)$$\\
 F^I_{rt} = \pa_r \left( e^g (X^I + \bar X^I) \right) ;\qquad
 \tilde F^I_{rt} = e^{2 g}\pa_r \left( e^{-g} (X^I - \bar X^I) \right)
 \end{array} \label{attraeqns}
\ee

We are interested in the near-horizon limit of a solution carrying
$q_0$ and $p^1$ charge. The requirement that $X^0$ is real imposes $q_0 p^1 <0$, and
in the following we will take $q_0<0,\  p^1>0$. The constants $h^I,\ h_I$ are related to
the asymptotic values of the K\"ahler moduli of $K_3 \times T^2$  as $r  \to \infty$.
Imposing the asymptotic condition of a regular 10-dimensional geometry $M^4 \times K_3 \times T^2$ as $r  \to \infty$
implies that the constants $h^I,\ h_I$ cannot all be put to zero. In our case,
we are required to take $h^A$ and $h_0$ to be nonzero, where $h^1 >0$ and where the form $h^i \o_i$ should
lie in the K\"ahler cone of $K_3$.

With  these asymptotic conditions one finds that in the
near-horizon $r \to 0$ region, the solution to (\ref{attraeqns}) reduces
to:
\be
\begin{array}{c}
 e^g \simeq r; \qquad
 X^0 \simeq - \sqrt{ p^1 \over |q_0|} ;\qquad
 X^1 \simeq {i p^1 \over 2} ;\qquad
 X^i \simeq {i h^i \over 2} r \\
 \hat{A} \simeq - 64; \qquad F^0 \simeq  2 \sqrt{ p^1 \over |q_0|} dt\wedge dr; \qquad
F^1 \simeq p^1 \sin \theta d \theta \wedge d \f
\end{array} \label{nearhor4d}
\ee
where $(\theta, \f)$ are coordinates on $S^2$.
The near-horizon metric is $AdS_2 \times S^2$. As in the discussion of \cite{Sen:2004dp}, we expect that the full
solution of (\ref{attraeqns}) which in the near-horizon limit reduces to (\ref{nearhor4d}) does not have the behavior
of flat Minkowski space with constant moduli at $r \to \infty$, but will rather contain unphysical fluctuations around it.
This is an artifact of a `bad' choice of field variables which should be removeable by making a suitable field redefinition
\cite{Sen:2004dp,Sen:2005kj}.

Let us briefly discuss the supersymmetry preserved by the
background (\ref{nearhor4d}). It should preserve at least 8
supersymmetries, since our charge configuration is
$1/2$-BPS. In the near-horizon limit, some enhancement may occur, and a
 maximally supersymmetric background would have 16 preserved supercharges.
 However, this cannot be the case here since the solution does not even preserve the
 full subset of $N=2$ supercharges. This would require all moduli to be constant \cite{LopesCardoso:2000qm}
 while the scalars $X^i$ in (\ref{nearhor4d}) vary linearly with $r$.
Therefore the preserved number of supersymmetries should be less
than 16, and more than or equal to 8. Curiously, the D0 brane
quantum mechanics we will write down in the next subsection will
have 12 super(conformal)-charges. Presumably, this number could be understood from
a detailed analysis of the $N=4$ supersymmetry variations which we shall not attempt here.

The  10-dimensional type IIA background metric and RR fluxes corresponding to (\ref{nearhor4d})
are given by
\be
\begin{array}{c}
ds^2 = Q^2 \left( - r^2 {dt^2} + {dr^2 \over r^2}  + d \theta^2 + \sin^2 \theta d \f^2
\right) + 2 dz d \bar z + 2 r g_{a \bar b} d z^a d \bar z^{\bar b}  \\ \  \\
g_{a \bar b} = {i g_s  \over 4 \p Q}   h^i (\o_i)_{a \bar b}; \qquad
F^{(4)} =  {p^1 \over 4 \p} \sin \theta d \theta d \f \wedge  \o_1; \qquad F^{(2)} =  {Q \over
g_s} dr \wedge dt
\end{array} \label{nearhor10d}
\ee
Here, we have chosen coordinates $(z, \bar z)$ on $T^2$ and
$(z^a, \bar z^{\bar a})_{a, \bar a=1,2} $ on $K_3$.
We will work in units in which $2 \p \sqrt{\a '} =1$. The radius
$Q$ of $AdS_2 \times S^2$ is then given by
$$
Q = {g_s \over 2 \p} \sqrt{ p^1 \over |q_0|}.
$$
An important difference of the background (\ref{nearhor10d}) with the large black hole backgrounds of \cite{Gaiotto:2004ij}
is that, in the latter case,
the magnetic $F^{(4)}$ flux permeates every 2-cycle on the compactification manifold,
while here it  only has a $T^2$ component. The $K_3$ is not supported by flux,
which is why its K\"ahler moduli are not fixed to finite values at the horizon
but vary linearly with $r$.

\subsection{Superconformal D0-brane quantum mechanics}

Following \cite{Gaiotto:2004pc,Gaiotto:2004ij}, we will consider the quantum mechanics of a nonabelian
configuration of $N$  D0-brane probes
in the background (\ref{nearhor10d}) corresponding to a  D2-brane wrapping the horizon
$S^2$.
This system has an alternative description in terms
of a horizon-wrapping D2-brane with $N$ units of flux turned on on its worldvolume \cite{Simons:2004nm,Gaiotto:2004pc}.
The bosonic part of the quantum mechanics can easily be derived from the
DBI and Wess-Zumino actions for the D2-brane, fixing a static gauge for the worldvolume reparametrizations.
The target space seen by the brane is $R\times T^2 \times K_3$
with metric
$$ ds^2 = T \left( 2 Q d \x^2 + {\x^2 \over Q} dz d\bar z + {1 \over Q} g_{a \bar b} d z^a d \bar z^{\bar b}\right) $$
where we defined $\x \equiv 1/\sqrt{r}$ and $T$ is the mass of a horizon-wrapped D2-brane with N units of flux:
$$ T = {2 \p \over g_s} \sqrt{(4 \p Q^2)^2 + N^2}.$$
Note that the target space is in this case a direct product $R\times T^2$ and $K_3$. This is a consequence of the
fact that the $K_3$ is not supported by flux and that its K\"ahler moduli vary linearly with $r$.

The bosonic Hamiltonian, to quadratic order in derivatives and in the limit $N \gg Q$, reads:
\bea
H_{ bos} &=& H^{\ R \times T^2}_{ bos} + H^{K3}_{ bos}\nonu
H^{R \times T^2}_{ bos} &=& {1\over 8QT} P_\x^2 + {Q \over T \x^2} (P_z - A_z)(P_{\bar z} -
A_{\bar z}) + {32 \p^4 Q^5 \over g_s^2 T \x^2}\nonu
H^{K3}_{ bos} &=&{Q \over T} P_a g^{a \bar b} P_{\bar b}
 \label{bosham}
\eea
The coordinates $(z^a, \bar z^{\bar a})$ on $K_3$ have been chosen such that the determinant of the $K_3$ metric
is constant.
We have introduced a $U(1)$ gauge potential $A$ on $T^2$ obeying
$ d A = 2 \p p^1 \o_1$. Explicitly, $A$ is given by\footnote{Note that our choice of coordinates on $T^2$ in (\ref{nearhor10d})
implies the normalization   $\o_1 = - {2 i \over \sqrt{|q_0| p^1 }} dz \wedge d \bar z$.}
\be A = { 4 \p^2 i Q \over g_s} \left( \bar z dz - z d \bar z \right) \label{Aeq} \ee
The dynamics on $R \times T^2$ and $K3$ decouples, a fact which, as we will see, will still be true when including
fermions.
This means that the symmetry group acting on the quantum mechanics
will naturally split into a product $G_1 \times G_2$ with $G_1$ and $G_2$ acting on the $R\times T^2$ and $K_3$
parts of the wavefunction
respectively. We shall show that $G_1$ is the $N=4$ superconformal group $SU(1,1|2)_Z$ (where $Z$ indicates
the presence of a central charge) and $G_2$ is the supergroup of $N=4$ supersymmetric quantum mechanics (SQM).
We shall now include the fermions and give the explicit form of the symmetry generators.

The kappa-symmetric action for a D2-brane in an arbitrary background \cite{Bergshoeff:1996tu} contains two
sixteen-component spinors
of $SO(9,1)$,
one of which can be eliminated by fixing kappa-symmetry. Hence the quantum mechanics contains sixteen fermions,
which are labeled as $(\l_\a, \bar \l_\a; \h_\a, \bar \h_\a; \h^a_\a, \h^{\bar a}_\a)_{\a = 1,2}$. These are roughly
the superpartners of
the bosonic coordinates $(\x;z ,\bar z;z^a, \bar z^{\bar a})$. The doublet index $\a$
indicates transformation properties under an $SU(2)$ R-symmetry which corresponds to spatial rotations.
The canonical anticommutation relations for the fermions are
\be
 \{ \l_\a,\bar \l_\b \} = \e_{\a \b};\qquad
\{ \h_\a,\bar \h_\b \} = \e_{\a \b};\qquad
 \{ \h^a_\a,\bar \h^{\bar b}_\b \} = \e_{\a \b} g^{a \bar b} \label{fermcomm}
 \ee

 The fermionic generators of the group $SU(1,1|2)_Z$ acting on the $R\times T^2$
 Hilbert space consist of supersymmetry generators $Q_\a, \bar
 Q_a$ and special supersymmetry generators $S_\a, \bar S_\a$ given
 by\footnote{Our conventions for $SU(2)$ index operations are "southwest-northeast", i.e. $\l^\a = \l_\b \e^{\b\a},\
 \l_\a =  \e_{\a\b}\l^\b,\ \l^2 = \l_\a\l^\a$ and we take $\e_{01} = \e^{01} =1$}:
 \bea
 Q_\a &=& {1 \over \sqrt{QT}} \left( \half \l_\a P_\x - {i \over \x} \h_{(\a} \bar
 \h_{\b)} \l^\b + {i \over 4 \x} \bar \l_\a \l^2 + {i \over 4 \x} \l_\a\right)\nonu
 &&+ \sqrt{Q \over T} \left( {\sqrt{2} \over \x} \h_\a (P_z - A_z) - {8 \p^2 Q^2 \over g_s} {i \over \x }\l_\a
 \right)\nonu
 \bar Q_\a &=& {1 \over \sqrt{QT}} \left( \half \bar \l_\a P_\x - {i \over \x} \h_{(\a} \bar
 \h_{\b)} \bar \l^\b - {i \over 4 \x} \bar \l^2  \l_\a  - {i \over 4 \x} \bar \l_\a\right)\nonu
 &&+ \sqrt{Q \over T} \left( {\sqrt{2} \over \x} \bar \h_\a (P_{\bar z} - A_{\bar z}) + {8 \p^2 Q^2 \over g_s} {i \over \x }\bar \l_\a
 \right)\nonu
 S_\a &=& 2 \sqrt{Q T} \x \l_\a\nonu
 \bar S_\a &=& 2 \sqrt{Q T} \x \bar \l_\a
\eea
In addition to the Hamiltonian $H^{R \times T^2}$, the bosonic generators of $SU(1,1|2)_Z$ consist of
the dilatation generator $D$, the generator of special conformal transformations $K$ and
the $SU(2)$ R-symmetry generators $T_{\a\b}$. They are given by
\be
\begin{array}{ccc}
D = \half (\x P_\x + P_\x \x );&\qquad &
K = 2 Q T \x^2 \\
T_{\a\b} = L^\l_{\a\b} +  L^\h_{\a\b};&\qquad& L^\l_{\a\b} = \l_{(\a} \bar \l_{\b)};\ L^\h_{\a\b}= \h_{(\a} \bar  \h_{\b)}
\end{array}\label{bosconfgen}
\ee
The relevant anticommutation relations are
\be
\begin{array}{ccc}
\{ Q_\a, \bar Q_\b \} = 2 \e_{\a\b} H^{R \times T^2};&\qquad& \{ Q_\a,  Q_\b \} =0   \\
\{ S_\a, \bar S_\b \} = 2 \e_{\a\b} K ;&\qquad& \{ S_\a,  S_\b \} =0 \\
\{ Q_\a, \bar S_\b \} =  \e_{\a\b} (D - {16 \p^2 i Q^3\over g_s}) - 2 i T_{\a\b}; &&
\{ S_\a, \bar Q_\b \} =  \e_{\a\b} (D + {16 \p^2 i Q^3\over g_s}) + 2 i T_{\a\b}
\end{array} \label{fermconfgen}
\ee

The $N=4$ SQM acting on the $K_3$ Hilbert space has  supersymmetry generators
$\cq_\a, \bar \cq_\a$ given by
\bea
\cq_\a &=& \sqrt{ 2 Q \over T}\h_\a^a P_a  \nonu
\bar \cq_\a &=& \sqrt{ 2 Q \over T}\bar \h_\a^{\bar a} P_{\bar a} \label{susygen}
\eea
with anticommutation relations
$$ \{ \cq_\a, \bar \cq_\b \} = 2 \e_{\a\b} H^{K_3}; \qquad \{ \cq_\a, \cq_\b \} = 0 .$$

\subsection{Counting chiral primaries and black hole entropy}\label{confcount}

Due to the existence of a dilatation generator $D$, the Hamiltonian $H$
(generating translations of Poincar\'e time) has a continuous spectrum, making the
counting of its ground states ill-defined. It was therefore proposed in \cite{Gaiotto:2004ij} to count instead
the ground states of $L_0 =  H + K$, the generator of global time translations.
From (\ref{bosham}, \ref{bosconfgen}) we see that $L_0$ has a bound state potential and its discrete
eigenstates will be localized in the radial $\x$ direction. The GSY proposal made in
\cite{Gaiotto:2004ij} states that the chiral primaries of the near-horizon D0-brane quantum mechanics
are to be identified with the black hole microstates. Applied to the case at hand, this means
that we have to count states of the form
$$|\psi \rangle \otimes |h\rangle$$
where  $|\psi \rangle$ is a chiral primary of
$SU(1,1|2)_Z$ and $|h\rangle$ is a supersymmetric
ground state of the $N=4$ SQM.

\subsubsection{Chiral primaries of $SU(1,1|2)_Z$}
Chiral primaries of $SU(1,1|2)_Z$ are characterized as follows \cite{Gaiotto:2004pc}.
We introduce the doublet notation
$$ Q^{++} = Q_1, \qquad Q^{-+} = Q_2, \qquad Q^{+-} = \bar Q_1, \qquad Q^{--} = \bar Q_2$$
and define
$$ G^{\a A}_{\pm \half} = {1 \over \sqrt{2}} ( Q^{\a A} \mp i S^{\a A} )$$
where $\a, A = +, -$. The anticommutation relations (\ref{fermconfgen}) now become
\bea
\{ G^{\a A}_{\pm \half}, G^{\b B}_{\pm \half} \}&=& \e^{\a\b} \e^{AB} L_{\pm 1}\nonu
\{ G^{\a A}_{ \half}, G^{\b B}_{-\half} \}&=& \e^{\a\b} \e^{AB} L_0 + 2 \e^{AB} T^{\a\b} + \e^{\a\b} Z^{AB}
\eea
where $Z^{AB}$ is a c-number central charge matrix with $Z^{++} = Z^{--} = 0,\ Z^{+-} = Z^{-+} = 16 \p^2 Q^3/ g_s >0$.
The second anticommutator implies a unitarity bound
\be L_0 \geq j + 16 \p^2 Q^3/ g_s \label{bound}\ee
with $j$ the spin under the $SU(2)$ R-symmetry.
Primary states are annihilated by the positive moded operators  $G_{\half}^{\a A}$.
Chiral primaries $\psi \rangle$ in addition saturate the bound (\ref{bound}),  hence they are also annihilated
by $G_{- \half}^{++}$:
\be G_{\half}^{\a A} | \psi \rangle = G_{- \half}^{++} | \psi \rangle =0. \ee

To construct the chiral primaries we use separation of variables into an $AdS_2$ and a $T^2$ component.
Denoting the $T^2$ component by $|\f\rangle$, we shall see that chiral primaries are in one-to-one
correspondence with states $|\f\rangle$ satisfying
\be
\h_\a (P_z - A_z)|\f\rangle = \bar \h_\a (P_{\bar z} - A_{\bar z})|\f\rangle = 0 \label{cond}
\ee
For the gauge field $A$ given in (\ref{Aeq}), the equation $P_{\bar z} - A_{\bar z} = 0$ has no
normalizeable solutions, while the solutions to $P_z - A_z = 0$ are  the lowest Landau level wavefunctions $\f_k(z, \bar z)$.
The number of independent
lowest Landau level wavefunctions is given by an index theorem and is equal to the first Chern number \cite{aharonov}
$${1 \over 2 \p}  \int_{T^2} d A = p^1.$$
Hence the equations (\ref{cond}) are solved by
\be
|\f_k\rangle = \f_k(z,\bar z) |0\rangle \label{T2sol}
\ee
where $|0\rangle$ is the vacuum state annihilated by the $\bar \h_\a$.
These $p^1$ states  are bosons under the $SU(2)$ of spatial rotations.

The construction of chiral primaries from the states $| \f_k \rangle$ now proceeds as follows.
On states obtained by tensoring $| \f_k \rangle$  with an arbitrary state in the  $AdS_2$ part of the Hilbert space,
the superconformal generators  $G^{\a A}_{\pm \half},\ G^{++}_{-\half}$ act as
\bea
G^{++}_{\pm \half} &=& {\l^{++} \over \sqrt{2 QT}} \left( \half  P_\x   + {i  \over 2 \x} \l^{+-} \l^{-+}
+ ( {1\over 4} - {8 \p^2 \over g_s} Q^3 ) {i \over \x} \mp 2 i QT \x \right)\nonu
G^{-+}_{ \half} &=& {\l^{-+} \over \sqrt{2 QT}} \left( \half  P_\x   - {i  \over 2 \x} \l^{--} \l^{++}
+ ( {1\over 4} - {8 \p^2 \over g_s}  Q^3 ) {i \over \x} - 2 i QT \x \right)\nonu
G^{+-}_{ \half} &=& {\l^{+-} \over \sqrt{2 QT}} \left( \half  P_\x   + {i  \over 2 \x} \l^{--} \l^{++}
- ( {1\over 4} - {8 \p^2 \over g_s}  Q^3 ) {i \over \x} - 2 i QT \x \right)\nonu
G^{--}_{ \half} &=& {\l^{--} \over \sqrt{2 QT}} \left( \half  P_\x   - {i  \over 2 \x} \l^{+-} \l^{-+}
- ( {1\over 4} - {8 \p^2 \over g_s}  Q^3 ) {i \over \x} - 2 i QT \x \right)
\eea
Normalizeable states annihilated by $G^{++}_{- \half}$ have to be annihilated by $\l^{++}$. Such states
automatically also have  $G^{++}_{ \half}=0$. If we choose the states to be annihilated
by $\l^{-+}$ as well, they will  have  $G^{-+}_{ \half}=0$ while the equations $G^{\pm-}_{ \half} = 0$
lead to a single differential equation for the $\x$-part of the wavefunction\footnote{Equivalently, one could proceed as
in \cite{Gaiotto:2004ij} and start instead from the vacuum annihilated by $\l^{++}$ and $\l^{+-}$. Adjusting the wavefunction
one can construct states that are annihilated by $G^{++}_{\pm \half}, G^{\pm-}_\half$ but not by $G^{-+}_\half$. Acting
with $G^{-+}_\half$ on these states one obtains chiral primaries. This construction leads to the same states (\ref{primaries}).}.
The resulting chiral primary states
are given by
\be
|\psi_k\rangle = \x^{- \half + {16 \p^2 Q^3 \over g_s}} e^{-2 QT \x^2} |\bar 0 \rangle \otimes |\f_k \rangle \label{primaries}
\ee
where $|\bar 0 \rangle$ is annihilated by $\l^{++}$ and $\l^{-+}$ and hence is bosonic under the rotational $SU(2)$. We have constructed
 in this manner $p^1$ bosonic chiral primary states of \linebreak $SU(1,1|2)_Z$.
 It's also possible to show that with the states (\ref{primaries}) we have found all chiral primaries.

\subsubsection{N=4 supersymmetric ground states}
A supersymmetric ground state in the $N=4$ supersymmetric quantum mechanics  satisfies
$$ \cq_\a | h\rangle  = \bar \cq_\a |h\rangle =0 .$$
Such states are well-known to be in one-to-one correspondence with the
Dolbeault cohomology classes on $K_3$ (see e.g. \cite{Denef:2002ru}). This can be seen
by representing states $| h\rangle $ by differential forms on $K_3$ and identifying
\bea
\h^a_1 \leftrightarrow d z^a, &\qquad& \bar \h_1^{\bar a} \leftrightarrow - d \bar z^{\bar b}\nonu
\h_2^{ a} \leftrightarrow g^{a \bar b} {\d \over \d( d \bar z^{\bar b})}, &\qquad&
 \bar \h_2^{\bar a} \leftrightarrow g^{\bar a  b} {\d \over \d( d  z^b)}
 \eea
 so that the anticommutation relations (\ref{fermcomm}) are satisfied.
Under this identification, bosonic and fermionic states are represented by even and odd forms respectively.
The supersymmetry generators are (up to  proportionality constants) identified with the Dolbeault operators
\bea
\cq_1 \leftrightarrow \pa &\qquad & \bar \cq_1 \leftrightarrow \bar \pa\nonu
\cq_2 \leftrightarrow \bar \pa^\dagger &\qquad& \bar \cq_2 \leftrightarrow  \pa^\dagger.
\eea
Since $K_3$ has 24 even harmonic forms, we find $24$ bosonic ground states of the $N=4$ SQM.

\subsubsection{Black hole entropy}
Tensoring together the chiral primaries of $SU(1,1|2)_Z$ and the supersymmetric ground states of
the N=4 SQM we find in total $24 p^1$ bosonic chiral primaries. Note that this number does
not depend on the background D0-brane charge $|q_0|$; hence for the purpose of counting ground states
we can take $q_0 \rightarrow 0$ and count the number of chiral primaries with total D0-brane charge $N$
in a background with fixed magnetic D4-charge $p^1$.
The large degeneracy of such states comes from the many ways the total number $N$ of D0-branes can be split into $k$ smaller
clusters of $n_i$ D0's such that
$$ \sum_{i = 1}^k n_i = N,$$
each cluster corresponding to a wrapped D2-brane that can reside in any of the $24 p^1$ bosonic chiral primary states.
The counting problem is the same as the counting the degeneracy $d_N$ of states at level $N$ in a $1+1$ dimensional CFT with $24 p^1$ bosons.
The generating function is then
\bea
Z &\equiv& \sum d_n q^n\nonu
&=& {\rm Tr}\; q^N\nonu
&=& \prod_n ( 1- q^n)^{-24 p^1}.\label{partK3}
\eea
This gives the asymptotic degeneracy at large $N$
$$
\ln d_N \approx 4 \p \sqrt{N p^1}
$$
which indeed corresponds to the known entropy of the black hole obtained either from microscopic
computation \cite{Dabholkar:1989jt,Dabholkar:1990yf,Dabholkar:1995nc} or from the supergravity description incorporating
higher derivative corrections \cite{LopesCardoso:1998wt,Dabholkar:2004yr,Dabholkar:2004dq,Dabholkar:2005by,Sen:2004dp,Hubeny:2004ji,Bak:2005mt,Sen:2005pu}.

\section{Quantum mechanics of the 2-charge black hole on $T^6$}\label{T6}
We will now consider a compactification on $T^6$ with the same charge configuration as before,
i.e. a background produced by $q_0$ D0-branes and $p^1$ D4-branes wrapping a $T^4$. These are $1/4$
BPS black holes of 4-dimensional $N=8$ supergravity and hence preserve the same number of supersymmetries as
the 2-charge black hole on $T^2 \times K_3$.  These black holes also have a vanishing horizon area
in the leading approximation, and the arguments of \cite{Sen:1997is} show that, in this case also, a horizon
is generated when including higher derivative corrections.

The situation is however less clear-cut than before,
as the  corrections to the prepotential vanish in this case.
Hence the corrections that generate the horizon are expected to come from non-holomorphic corrections
to the supergravity equations, and it is not known how to incorporate these systematically at present. In this section,
we shall be a little cavalier  and simply  assume that the near-horizon limit of the quantum corrected background
is still of the form (\ref{nearhor10d}), with the $K_3$ metric now replaced by the  metric on $T^4$ and possibly with a different
value of the constant $Q$.

The resulting
quantum mechanics of a wrapped D2-brane with D0-brane flux then reduces to (\ref{bosham},\ref{fermconfgen},\ref{susygen})
with $g_{a \bar b}$ replaced by the flat
metric on $T^4$. The symmetry group again splits into a superconformal $SU(1,1|2)_Z$ acting on the
$R \times T^2$ part of the Hilbert space and an $N=4$ supersymmetric quantum mechanics  acting on the
$T^4$ part. The counting of chiral primaries of $SU(1,1|2)_Z$ goes trough as in section \ref{confcount}, yielding
$p^1$ bosonic chiral primaries. The difference lies in the counting of ground states  of the
$N=4$ SQM, now corresponding to the Dolbeault cohomology of $T^4$. Since $T^4$ has 8 even and 8 odd harmonic forms,
we find   8 bosonic and 8 fermionic ground states. The counting problem for chiral primaries is now isomorphic to
counting the degeneracy at level $N$ of a CFT with $8 p^1$ bosons and $8p^1$ fermions. The partition function is
\be Z = \prod_n \left( {1 + q^n \over 1 - q^n} \right)^{8 p^1}. \label{partT6}\ee
This gives the asymptotic degeneracy
$$ \ln d_N \approx 2 \sqrt{2} \p \sqrt{N p^1} $$
which is in agreement with the known degeneracy from microscopic counting \cite{Dabholkar:1989jt,Dabholkar:1990yf,Dabholkar:1995nc}.

\section{Discussion}

In this paper we provided additional evidence for the GSY proposal by counting the asymptotic degeneracy
of chiral primaries in the superconformal quantum mechanics describing the near horizon physics of small black holes
and showing it to agree with the black hole entropy.
We now list some open questions.
\begin{itemize}
\item Our construction of the quantum mechanical superconformal symmetry algebra  suggests
that the number of supercharges in the near-horizon background (\ref{nearhor4d}) is enhanced from 8 in the
bulk to 12. It would be interesting to prove this directly from  the $N=4$ supersymmetry variations.
\item In section \ref{T6} we assumed  the near-horizon geometry of the 2-charge black hole on $T^6$ including quantum corrections
to be of the form (\ref{nearhor10d}), and saw that this led to the correct counting of chiral primaries. The quantum corrections
are expected to come from non-holomorphic
corrections to the supergravity equations and it would be of interest to check whether known nonholomorphic corrections
such as $R^4$ terms indeed lead to  (\ref{nearhor10d}).
\item It is not clear in how far the agreement found in this work depended on the large number of 8 supersymmetries
preserved by the small black holes considered here. In particular, it would be interesting to verify whether for small black holes
which are 1/2 BPS states in an $N=2$ compactification, preserving only 4 supersymmetries, the counting of chiral primaries in the
D0-D2 quantum mechanics still reproduces the correct entropy
formula.
\item The small black holes considered here could also prove to be a good testing
ground for verifying or refining the GSY proposal in order to  reproduce the correct subleading corrections to the entropy formula.
  It should be noted that, although the chiral primary partition functions (\ref{partK3}, \ref{partT6}) only reproduce the
leading term in the entropy formula, a  modification of the counting problem would produce  the microcanonical partition function
to all orders in both cases. Whether such a modification can be justified in this context remains to be seen.
\end{itemize}

\acknowledgments{We would like to thank Piljin Yi for a useful discussion and Davide Gaiotto for correspondence.}

\end{document}